\documentclass[11pt,showpacs,showkeys]{revtex4}
\input epsf.sty

\usepackage{ulem}
\usepackage{cancel}
\usepackage{color}

\def\red#1{\textcolor{red}{#1}}

\def\comment#1{}

\usepackage{graphicx}
\begin{document}

\title{Cosmological constant, matter, cosmic inflation and coincidence}
\author{She-Sheng Xue}
\email{xue@icra.it; shesheng.xue@gmail.com}
\affiliation{ICRANet, Piazzale della Repubblica, 10-65122, Pescara,\\
Physics Department, Sapienza University of Rome, P.le A. Moro 5, 00185, Rome,
Italy} 


\begin{abstract}
We present a possible understanding to the issues of cosmological constant, 
inflation, matter and coincidence problems based only on the Einstein 
equation and Hawking particle production. The inflation appears and 
results agree to observations. The CMB large-scale anomaly can be explained and the dark-matter acoustic wave is speculated. The entropy and reheating are discussed. The cosmological term $\Omega_{_\Lambda}$ tracks down the matter $\Omega_{_M}$ until the radiation-matter equilibrium, then slowly varies, 
thus the cosmic coincidence problem can be avoided. 
The relation between $\Omega_{_\Lambda}$ and $\Omega_{_M}$ is shown and can be examined at large redshifts.       
\end{abstract}



\maketitle

\noindent
{\bf Introduction.}
In the standard model of modern cosmology ($\Lambda$CDM), the cosmological constant, inflation, dark matter and coincidence problem have been long standing issues since decades, though many models and efforts have been made to approach these issues, 
and readers are referred to 
review articles and professional books, for example, 
see Refs.~\cite{Peebles,kolb,book,Inflation_R,
Inflation_higgs,reviewL,Bamba:2012cp,Nojiri:2017ncd,
Coley2019,Biagetti2019,Sola2019}. We present here the possible scenario based only on the Einstein equation, in which the cosmological term 
generates (couples to) the matter via the Hawking pair production of particles and antiparticles.  
As an effective field theory of the Einstein gravity,
two physically relevant area operators of the Ricci scalar 
$R$ and cosmological term $\Lambda$ can be possibly realized in the scaling domain \cite{w2}, 
\begin{eqnarray}
{\mathcal A}^{^{\rm eff}}_{\rm EC}
&=&\int \frac{d^4x}{16\pi G}{\rm det}(-g)(R-2\Lambda),
\label{ec0}
\end{eqnarray}
and high-dimensional operators are suppressed. The gravitation constant $G\sim \ell^2_{\rm pl}=M^{-2}_{\rm pl}$ 
is the smallest area at the Planck cutoff. Whereas the cosmological constant represents the intrinsic scale 
$\Lambda\propto \xi^{-2}$, 
the scaling invariant correlation length square 
$\xi^2$ is the largest area at the Universe  
horizon \cite{cosh}. We will further show such a dynamics of cosmological constant $\Lambda$ and its function in Universe evolution.


The effective action (\ref{ec0}) yields the Einstein equation for the spacetime of Einstein tensor 
${\mathcal G}^{ab}$ coupling to the matter of energy-momentum 
tensor $T^{ab}_{_M}$,
\begin{eqnarray}
{\mathcal G}^{ab} = -8\pi G T^{ab}_{_M}; \quad
{\mathcal G}^{ab} =  R^{ab} -(1/2) g^{ab}R -\Lambda g^{ab}.
\label{e1}
\end{eqnarray}   
Its covariant differentiation and the Bianchi identity are 
\begin{eqnarray}
{\mathcal G}^{ab}_{\,\,\,\,\,\,;\,b} = -8\pi [\,G \,T_{_M}^{ab}\,]_{\,;\,b}, \quad
[ R^{ab}-(1/2)\delta^{ab}R]_{\,;\,b}\equiv 0,
\label{de1}
\end{eqnarray}
which lead to the conservation law,
\begin{equation}
(\Lambda)_{;\,b}\,g^{ab}=8\pi (G)_{;\,b}T^{ab}_{_M}+8\pi G 
(T^{ab}_{_M})_{;\,b}\,\, ,
\label{geqi00}
\end{equation} 
with varying cosmological term $(\Lambda)_{;\,b}=(\Lambda)_{,\,b}$ and  coupling $(G)_{;\,b}=(G)_{,\,b}$. Despite its {\it essence} of spacetime origin, the cosmological $\Lambda$-term 
in ${\mathcal G}_{ab}$ can be moved to the RHS of Eq.~(\ref{e1}) 
and {\it formally} expressed as $T^{ab}_{_{_{\Lambda}}}$, analogously to the $T^{ab}_{_{M}}$ of a perfect fluid,
\begin{eqnarray}
T^{ab}_{_{M,\,\Lambda}} &=&p_{_{_{M,\,\Lambda}}} g^{ab} +(p_{_{_{M,\,\Lambda}}}  + \rho_{_{_{M,\,\Lambda}}} ) U^aU^b,\label{emt}
\end{eqnarray} 
by implementing a negative mass density 
$\rho_{_\Lambda}=\Lambda/(8\pi G)\equiv -p_{_\Lambda}$. 
Equation (\ref{geqi00}) is equivalent to the conservation 
law $T^{ab}_{\,\,\,\,\,\,;\,b}\equiv (T^{ab}_{_M} + T^{ab}_{_\Lambda})_{;\,b}=0$.
The energy density $\rho_{_{_{M,\,\Lambda}}}$ and pressure $p_{_{_{M,\,\Lambda}}}$ are in the comoving frame of four 
velocity $U^a=(1, 0,0,0)$. 

In the Robertson-Walker spacetime $ds^2=a^2(t)d{\bf x}^2$ of zero curvature, 
Eqs.~(\ref{e1}) and (\ref{geqi00}) become,
\begin{eqnarray}
&&h^2 = g(\Omega_{_M} + \Omega_{_\Lambda}),~~ h=H/H_\circ, 
~~ \Omega_{_{M,\Lambda}}=\rho_{_{M,\Lambda}}/\rho_c^\circ, ~~~
\label{e3}\\
&&\frac{dh^2}{dx}\! +\! 2h^2 \!=\!
g\Big[2\Omega_{_\Lambda}\!-\!(1\!+\!3\omega_{_M})\Omega_{_M}\Big],
\label{e2}\\
&&\frac{d}{dx}\left[g(\Omega_{_\Lambda}+\Omega_{_M})\right]
=-3g(1+\omega_{_M})\Omega_{_M},
\label{cgeqi20}
\end{eqnarray}
where $g=G/M^{-2}_{\rm pl}$, $H=\dot a/a$,  
$\omega_{_M} = p_{_M}/\rho_{_M}$, $x=\ln(a/a_\circ)$, and $d(\cdot\cdot\cdot)/dt=Hd(\cdot\cdot\cdot)/dx$ \cite{xueNPB2015}.
The characteristic scales $H_\circ$, $a_\circ$, and $\rho^\circ_c=3H^2_\circ/(8\pi M^{-2}_{\rm pl})$ depend on the Universe evolution epoch: inflation, reheating, radiation and matter dominated epochs.

Here we consider only the constant $G
=M^{-2}_{\rm pl}$ \cite{gvary} 
and set the reduced Planck scale $8\pi G
=m^{-2}_{\rm pl}=1$, unless otherwise stated. 
Equations (\ref{e3},\ref{e2}) and (\ref{cgeqi20}) are recasted into two independent equations,
\begin{eqnarray}
&& h^2 = (\Omega_{_M} + \Omega_{_\Lambda}),
\label{fe3}\\
&&\frac{d}{dx}\left(\Omega_{_\Lambda}+\Omega_{_M}\right)
=-3(1+\omega_{_M})\Omega_{_M},
\label{fcgeqi20}
\end{eqnarray}
which reduces to the Friedmann equations for the constant cosmological 
term $\Omega_{_\Lambda}$. Equations (\ref{cgeqi20}) and (\ref{fcgeqi20}) show
the interaction between the cosmological term and matter. Moreover, the 
matter term $\Omega_{_M}(h)$ is generated by the spontaneous 
pair production of particles and antiparticles from the spacetime horizon $h$, as
will be shown in next section. 
In turn, $\Omega_{_M}(h)\not=0$ dynamically
leads to $h^2$ and $\Omega_{_\Lambda}$ decrease 
via Eq.~(\ref{fcgeqi20}), it changes via Eq.~(\ref{fe3}). 
This completely determines the variations of 
$h^2(x)$ and $\Omega_{_\Lambda}(h)$ and $\Omega_{_M}(h)$ scaling in the
Universe evolution. 



\noindent
{\bf Pair production.}
To calculate $\Omega_{_M}(h)$, we consider the pair production of 
spin-$1/2$ particles and antiparticles in the exact De Sitter spacetime 
of the constant $H$ and scaling factor $a(t)=e^{Ht}$. 
The averaged number density of pairs produced from 
$t_\circ=0$ to $t\approx 2\pi H^{-1}$ is \cite{lnt2014,ekhard}
\begin{eqnarray}
n_{_M} &=& \frac{H^3}{2\pi^2} \int_0^\infty dz z^2 |\beta^{(n)}_k(t)|^2\nonumber\\
&=&\frac{H^3e^{\pi\mu}}{16\pi} \int_0^\infty dz \frac{z^3}
{\sqrt{z^2+\mu^2}} {\mathcal F}^{(n)}_\nu(z,\mu),
\label{nden}
\end{eqnarray}
where $z\equiv kH^{-1}e^{-Ht}$, the particle mass $\mu=m/H$ and momentum $k$, the Bogolubov coefficient up to the $n$-th adiabatic order 
$|\beta^{(n)}_k(t)|_{k\rightarrow \infty}\sim {\mathcal O}(1/k^{n+2})$
in the ultraviolet (UV) limit. Due to the exact De Sitter symmetry 
$(H={\rm const})$, the energy-momentum tensor of produced pairs 
$T^{\mu\nu}_{_M}\propto g^{\mu\nu}$ \cite{mottola,lnt2014}. 
Since the back reaction of pair production leads to a slowly decreasing $H$ and breaks the exact symmetry, we assume
$T^{\mu\nu}_{_M}$ to be spatially homogenous and in 
the form (\ref{emt})   
\begin{eqnarray}
\rho_{_M} &=& 2\,\frac{H^3}{2\pi^2} \int_0^\infty dz z^2
\epsilon_k|\beta^{(n)}_k(t)|^2\nonumber\\
&=& 2\,\frac{H^4e^{\pi\mu}}{16\pi} \int_0^\infty dz z^3 {\mathcal F}^{(n)}_\nu(z,\mu) ,
\label{rden}\\
p_{_M} &=& 2\,\frac{H^3}{2\pi^2} \int_0^\infty dz z^2 \frac{(k/a)^2}{
3\epsilon_k}|\beta^{(n)}_k(t)|^2\nonumber\\
&=& \frac{\rho_{_M}}{3}
-2\,\frac{\mu^2H^4e^{\pi\mu}}{3\times 16\pi} \int_0^\infty dz \frac{z^3}
{z^2+\mu^2} {\mathcal F}^{(n)}_\nu(z,\mu),~~~~~
\label{pden}
\end{eqnarray} 
where the spectrum of created particles $\epsilon_k=a^{-1}[(k/a)^2+m^2]^{1/2}$. 
To ensure the UV finiteness of Eqs.~(\ref{nden}), (\ref{rden}) and (\ref{pden}), the appropriate adiabatic 
order $n$ is considered, 
\begin{eqnarray}
{\mathcal F}^{(n)}_\nu(z,\mu) &=&\Big|f_1^{(n)}\sigma_+H^{(1)}_{\nu-1}(z)-i f_2^{(n)}\sigma_-H^{(1)}_{\nu}(z)\Big|^2 ,~~~~~
\label{fden}
\end{eqnarray} 
where $\sigma_\pm\equiv [(z^2+\mu^2)^{1/2}\pm\mu]^{1/2}$, $\nu=1/2-i\mu$, $H^{(1)}_{\nu}(z)$ is the Hankel function of the first kind, 
and $f_{1,2}^{(n)}=1+\sum_{i=1}^n F_{1,2}^{(i)}$ 
\cite{lnt2014}.
 
The spacetime of the horizon $H$ produces particles and antiparticles 
of different masses $m \gtrsim H$ and degeneracies $g_d$. 
We simply introduce 
the unique mass scale ``$m$'' 
to effectively describe the total contribution of pairs 
to Eqs.~(\ref{nden}), (\ref{rden}) and (\ref{pden}), and 
its value is determined by observations. These particles and antiparticles 
can be both dark matter and usual matter particles. It should be also noted that  
the pair productions of bosonic particles and
antiparticles are not considered here, since their number density 
$n^B_{_M}$ goes to zero for $m/H\gg 1$ and has a spurious divergence 
for $m/H \ll 1$ \cite{lnt2014}. Their quantitative contributions 
to the energy density and pressure of matter content are postponed  
for future studies.

\begin{figure}   
\includegraphics[height=4.5cm,width=6.8cm]{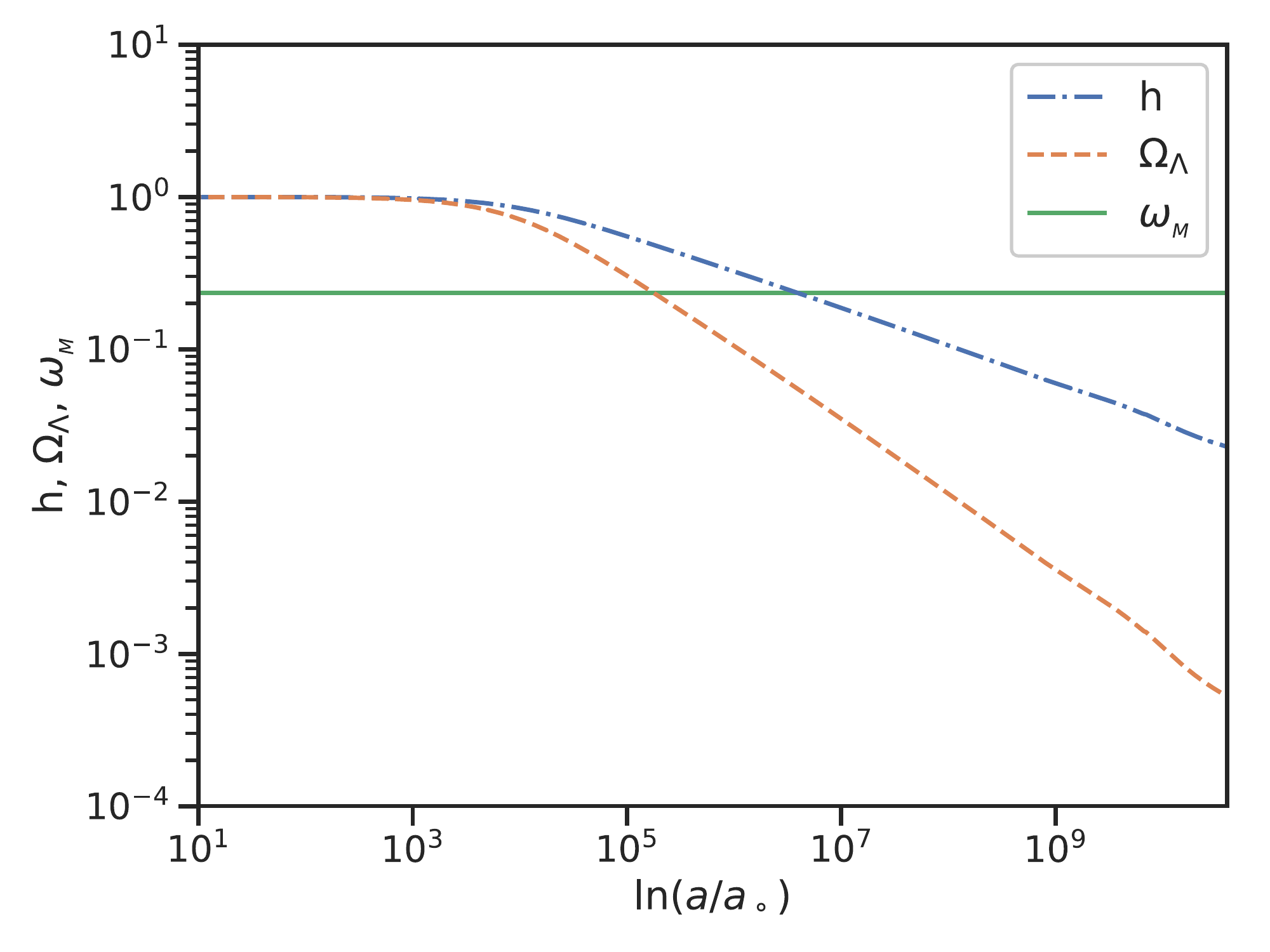}
\vspace{-1em}
\caption{The inflation appears, as
$h$ and $\Omega_{_\Lambda}(h)$, $\omega_{_M}=p_{_M}/\rho_{_M}$ 
slowly decrease 
in the $e$-folding number. Here $m=H_\circ=1$. Note that the superscript or subscript ``$\circ$'' indicates the quantities at the inflation beginning, not to be confused with ``$0$'' standing for the present time.}
\label{finflation}
\end{figure}

\noindent
{\bf Cosmic inflation.}
Starting from the initial conditions 
$\Omega^\circ_{_\Lambda}=h^2_\circ$  
at the reduced Planck scale $\Lambda_\circ=3H^2_\circ \sim m^2_{\rm pl}$, 
$\Omega_{_\Lambda}(h)\gg \Omega_{_M}(h)$, and $\Omega_{_\Lambda}(h)$ 
governs the varying spacetime horizon $h$ in the inflation epoch.
Here we select the initial scale 
$H_\circ=m_{\rm pl} < M_{\rm pl}$
so that the details of quantum gravity and/or Planck transition could possibly be ignored and Eqs.~(\ref{fe3}) 
and (\ref{fcgeqi20}) approximately hold. Numerically integrating Eqs.~(\ref{fe3},\ref{fcgeqi20}) and (\ref{rden},\ref{pden}) with the initial condition $h^2_\circ =1$ and 
$h^2_\circ\gtrsim\Omega^\circ_{_\Lambda}\gg \Omega^\circ_{_M}$,
we find that the cosmic inflation of very slowly decreasing $h^2$ and 
$\Omega_{_\Lambda}(h)$ is indeed a solution, as illustrated in 
Fig.~\ref{finflation}. The reason is that the pair production (\ref{nden}) is not so rapid that 
$h^2$ decreases slowly, see Eq.~(\ref{fcgeqi20}), as a function of $e$-folding numbers $\ln (a/a_\circ)$. This in turn justifies our approximate calculations (\ref{rden}) and (\ref{pden}) by using formulas for a constancy $H$. 
As a result, we obtain the solution to the 
cosmological ``constant'', slowly 
varying as an ``area'' law $\Lambda=3H^2_\circ\Omega_{_\Lambda}(h)\approx 3 H^2$ or $\Omega_{_\Lambda}(h)\approx h^2$.  

Due to the continuous pair productions, $\Omega_{_M}/\Omega_{_\Lambda}$ increasing, $H$ and $\Omega_{_\Lambda}$ decreasing, the 
inflation ends at $a=a_e$, which can be estimated by the expansion rate $H_e$ being smaller than the averaged pair-production rate $\Gamma_M \approx dN/(2\pi dt)\approx (H/2\pi )dN/dx$ and the number 
of particles $N=n_{_M}H^{-3}/2$. 
\comment{
\begin{eqnarray}
\Gamma_M \equiv d(n_{_M}V)/dt= d(n_{_M}V)/Hdx,
\label{prate}
\end{eqnarray}
where the volume $V=(4\pi/3) H^{-3}$. }
However, $H_e < \Gamma_M(H_e)$ 
occurs in $m \gg H$, 
where it is difficult to perform numerical calculations 
of Hankel functions \cite{num}. 

To explicitly show the 
inflation physics in $m \gg H$, we explore  asymptotic expressions: 
\begin{eqnarray}
n_{_M} &\approx & \chi m\, H^2,\quad \Gamma_M \approx - 
(\chi m/4\pi) (H^{-1}dH/dx),~~~
\label{aden}\\
\rho_{_M} &\approx & 2\,\chi m^2 H^2(1+ s
),
\quad p_{_M} \approx  (s/3) \rho_{_M},
\label{apden}
\end{eqnarray} 
$\omega_{_M} \approx   s/3$, where  
$\chi\approx 1.85\times 10^{-3}$ and $s\approx 1/2(H/m)^2$ 
\footnote{Ref.~\cite{lnt2014} shows that the number density 
(\ref{nden}) asymptotically approaches $n_{_M} \sim m H^2$. We numerically determine $\chi$ and insert the damping factor 
$e^{-\sigma(z^2+\mu^2)}$ into $n_{_M}\rightarrow n_{_M}(\sigma)$ to estimate the $s$-term by the saddle-point approximation.  
}.
The leading order of both $n_{_M}$ and $\rho_{_M}$ follows 
the area law $\propto H^2$, rather than the volume law (\ref{nden}-\ref{pden}).
The physical picture is the large number (or degeneracies $g_d$) $N\sim H^{-1}/m^{-1}\gg  1$ of pairs produced mainly in the thin layer of the width $1/m$ on the horizon surface area $H^{-2}$.    

Consequently, Eq.~(\ref{fcgeqi20}) becomes 
\begin{eqnarray}
dH^2/dx &\approx& - 2\,\chi \,m^2 H^2 (1+\omega_{_M})(1+s),\label{apps}
\end{eqnarray}
yielding $H \approx H_*\exp -\chi m^2x=H_*(a/a_*)^{-\chi m^2}$, 
slowly decreasing for $\chi m^2=\chi (m/m_{\rm pl})^2\ll 1$. This solution 
shows the features of the inflation epoch. The initial scale  
$H_*$ 
corresponds to 
the interested mode of the pivot scale $k_*$ crossed the horizon $(c_sk_*=H_*a_*)$ for CMB observations. At this pivot scale,
one calculates the scalar, tensor power spectra and their ratio 
\begin{eqnarray}
\Delta^2_{_{\mathcal R}} 
= \frac{1}{8\pi^2}\frac{H^2_*}{m^2_{\rm pl}\,\epsilon\,c_s};
\Delta^2_h 
= \frac{2}{\pi^2}\frac{H^2_*}{m^2_{\rm pl}};~ r\equiv \frac{\Delta^2_h}{\Delta^2_{_{\mathcal R}}}=16\,\epsilon\, c_s,~
\label{ps}
\end{eqnarray} 
where 
$c_s< 1$ due to the Lorentz symmetry broken by the time dependence of the background \cite{book}.
\comment{
including c_s the velocity in Lorentz violation vacuum 
\begin{eqnarray}
\Delta^2_{_{\mathcal R}} (k)
= \frac{1}{8\pi^2}\frac{H^2}{m^2_{\rm pl}\epsilon c_s^2},\quad 
\Delta^2_h (k) 
= \frac{2}{\pi^2}\frac{H^2}{m^2_{\rm pl}};\quad r\equiv \frac{\Delta^2_h (k)}{\Delta^2_{_{\mathcal R}} (k)}=16\epsilon c_s^2,
\label{ps}
\end{eqnarray}
and their deviations from the scale invariance 
$\Delta^{(n)}_{_{\mathcal R}} (k)\equiv d^n  \ln \Delta_{_{\mathcal R}} (k)/d (\ln k)^n|_{k_*}$ and $\Delta^{(n)}_h (k)\equiv d^n  \ln \Delta_h (k)/d (\ln k)^n|_{k_*}$:
including sound velocity $c_s, \kappa$
\begin{eqnarray}
n_s-1 &=& \Delta^{(1)}_{_{\mathcal R}} (k_*)
\approx -2\epsilon -\eta -\kappa,
\label{ns}\\
\alpha_s &=& \Delta^{(2)}_{_{\mathcal R}} (k_*)
 \approx  -(2\epsilon^\prime +\eta^\prime +\kappa^\prime)\approx n_s^\prime
\label{rs}\\
\tilde\alpha_s &=& \Delta^{(3)}_{_{\mathcal R}} (k_*)\approx -(2\epsilon^{\prime\prime} +\eta^{\prime\prime} +\kappa^{\prime\prime})\approx \alpha_s^\prime
\label{rrs}\\
n_t &=& \Delta^{(1)}_h (k_*)=-2\epsilon 
\label{nt}\\
\tilde n_t &=& \Delta^{(2)}_h (k_*) =  (1-\epsilon-\kappa)^{-1} d(-2\epsilon)/d \ln x \approx n_t^{\prime}.
\label{rnt}
\end{eqnarray} 
}
Their deviations from the scale invariance 
$\Delta^{(n)}_{{\mathcal R},h}\equiv d^n  \ln \Delta_{{\mathcal R},h} (k)/d (\ln k)^n|_{k_*}\approx d^n  \ln \Delta_{{\mathcal R},h} (k_*)/d x^n$:
\begin{eqnarray}
n_s-1 &=& \Delta^{(1)}_{_{\mathcal R}}
\approx -2\epsilon -\eta-\kappa,~~ \alpha_s = \Delta^{(2)}_{_{\mathcal R}}
\approx n_s^\prime ~~~
\label{ns}\\
n_t &=& \Delta^{(1)}_h =-2\epsilon, ~~\,\,\,\,\,\,\,\,\,\,\,\,\,\,\,~~~~
\tilde n_t = \Delta^{(2)}_h \approx n_t^{\prime},~~~ \label{nt}
\end{eqnarray} 
and $\tilde\alpha_s = \Delta^{(3)}_{_{\mathcal R}}
\approx \alpha_s^\prime 
$, and we calculate
\begin{eqnarray}
\epsilon &\equiv& -H'/H |_{k_*}
\approx \chi\, m^2(1+s),\label{ep}\\ 
\eta &\equiv& \epsilon'/\epsilon |_{k_*}
\approx -3\chi\, m^2s\approx -3\, s\,\epsilon,
\nonumber
\end{eqnarray} 
$\kappa = c\,'_s/c_s$ and their derivatives 
$\eta' = d \eta/ d x \approx -3\eta \epsilon^2$, $\epsilon''\approx \eta^2\epsilon-3\eta\epsilon^3$, $\eta'' \approx 9\eta\epsilon^4-6\eta^2\epsilon^2$. 

\comment{Due to the spontaneously breaking of De Sitter symmetry, Goldstone boson $\pi$ appears and $\kappa' = d \kappa/ dx \approx -4\kappa \epsilon^2$, as functions of $e$-folding numbers $\ln (a/a_\circ)$, where $y^\prime  \equiv d y/dx$ and $y^{\prime\prime}\equiv d^2 y/dx^2$.}
\comment{
and the observational values $n_s\approx 1-2\epsilon \approx 0.96$ and 
$2\epsilon \approx 0.04$ leading to $m =4.63$, i.e., $m=1.13\times 10^{19}
{\rm GeV} \lesssim M_{\rm pl}=1.22 \times 10^{19}$ GeV. If we put $m=M_{\rm pl}$, $\epsilon = 0.047$. Suppose that the inflation ends when the rate (\ref{prate}) is significantly larger than the expansion rate $H$, parameterizing as 
$\Gamma =(3/8\pi)\alpha H$ and $\alpha >(8\pi/3)$,} 

\begin{figure}   
\includegraphics[height=7.5cm,width=7.8cm]{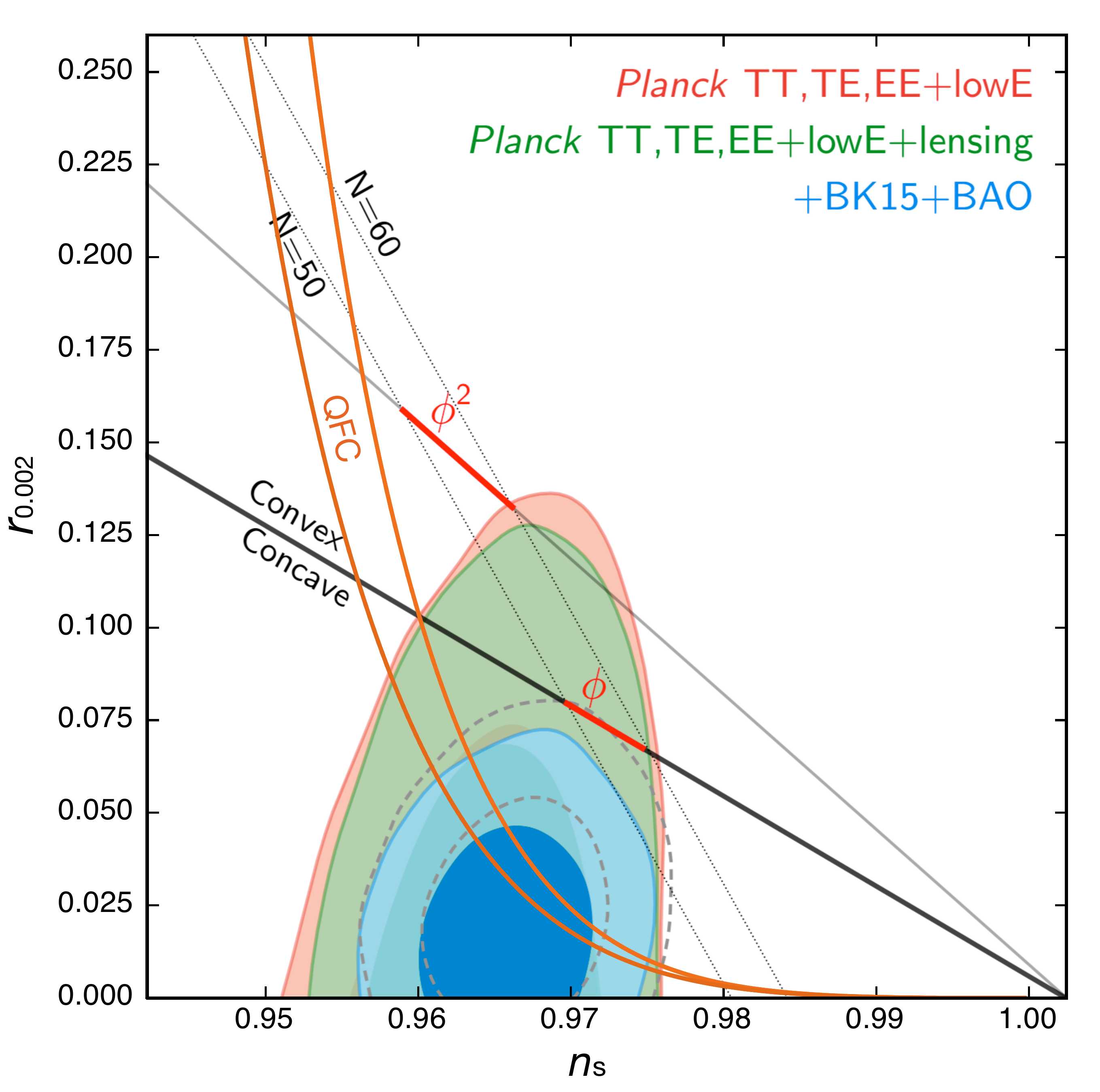}
\vspace{-1em}
\caption{On the Figure 28 of the Planck 2018 results \cite{Planck2018} for constraints on the tensor-to-scalar ratio $r$,  we plot 
the parameter-free ($n_s-r$) relation (\ref{endr}) that shows 
in the observed $n_s$-range, two QFC curves respectively representing $N_{\rm end}=60$ and $N_{\rm end}=50$ are consistently inside the blue zone constrained 
by several observational data sets. The real values of $r$ ratio should be below the curves
due to the nature of inequality (\ref{endr}). As a short notation, the abbreviation 
QFC stands for the name ``quantum field cosmology'' \cite{xueNPB2015,clement}.}
\label{ns-rplot}
\end{figure}

Based on two observational values 
at $k_*=0.\,05\, ({\rm Mpc})^{-1}$ \cite{Planck2018}: (i) 
$n_s\approx 0.965$ , 
we estimate 
$m \lesssim 3.08\, m_{\rm pl}$ by 
$2\epsilon\approx 2\chi  m^2\lesssim 1-n_s\approx 0.035$ 
for $\epsilon \gg \eta$ and assuming $2\epsilon < \kappa$; (ii)
$\Delta^2_{_{\mathcal R}}=A_s 
\approx  2.1\times 10^{-9}$, Eq.~(\ref{ps}) gives the inflation scale $H_*=3.15\times 10^{-5}\,(r/0.1)^{1/2}m_{\rm pl}$, and Eq.~(\ref{aden}) gives the pair-production rate $\Gamma^*_M=(\chi m/4\pi)\epsilon=7.9\times 10^{-6}m_{\rm pl}$.
The inflation ends at $\Gamma_M > H_e$, i.e., 
$(\chi m/4\pi)\,\epsilon > H_*\exp -(\epsilon\, N_e)$, 
\begin{eqnarray}
N_e=\ln \left(\frac{a_e}{a_*}\right)
> \frac{2}{1-n_s}\ln\left[\frac{7.91\cdot 10^{-4}\,(r/0.1)^{1/2}}{ (1-n_s) \, \chi \,(m/m_{\rm pl})}\right],~~~
\label{end}
\end{eqnarray}  
yielding the results $r <0.037,\, 0.052$ for $N_e=50,\,60$, 
in agreement with observations \cite{Planck2018}. Replacing the unique mass 
parameter $m$ by the observed quantity of spectral index $n_s$:  
$2\chi(m/m_{\rm pl})^2\approx (1-n_s)$. As a result, 
being independent of any free parameter, 
Equation (\ref{end}) yields a definite ($n_s-r$)-relationship between the spectral index $n_s$ 
and the scalar-tensor-ratio $r$, 
\begin{eqnarray}
(r/0.1)<7.97\times 10^{5}\chi (1-n_s)^{3} e^{(1-n_s)N_{\rm end}},
\label{endr}
\end{eqnarray}
for a given $N_{\rm end}$ value of 
inflation $e$-folding number, see Fig.~\ref{ns-rplot}. Moreover, $n'_s< \epsilon ^2\approx (1-n_s)^2/4$,
$n''_s < \epsilon ^3\approx (1-n_s)^3/8$, and we need to 
know $\kappa$ for further parameter constrains. 
In this inflation epoch $H_*>H>H_e$,  
$\Omega_{_\Lambda}= (H/H_*)^2-\Omega_{_M}$ dominates over 
$\Omega_{_M}\approx (\chi m^2/3)(H/H_*)^2$, and the cosmological ``constant'' $\Lambda= 3H_*^2\Omega_{_\Lambda} \propto H^2$. 

Using Eqs.~(\ref{fe3}) and (\ref{fcgeqi20}), we recast Eqs.~(\ref{ps}) 
and (\ref{ep}) as 
\begin{eqnarray}
\Delta^2_{_{\mathcal R}} (k) 
= \frac{1}{12\pi^2}\frac{H^2{\mathcal R}^{-1}_{_M}}{m^2_{\rm pl}(1+\omega_{_M})c_s},~~ \epsilon = \frac{3}{2}(1+\omega_{_M}){\mathcal R}_{_M},~~
\label{ps1}
\end{eqnarray}
where ${\mathcal R}_{_M}=\Omega_{_M}/(\Omega_{_\Lambda}+\Omega_{_M})$.
In the ``pre-inflation'' epoch $H_\circ > H>H_*$, see Fig.~\ref{finflation},
$\omega_{_M}$ varies from 
$\sim 1/3$ ($H/m \sim 1$) to $0$ ($H/m\ll 1$), while
$H$ and $\Omega_{_\Lambda,_M}$ 
slowly vary a few percent only, 
implying that $\Delta^2_{_{\mathcal R}} (k)$ (\ref{ps1}) 
decreases $3/4$ at most. 
This probably explains the large-scale anomaly of
the low amplitude of the CMB power spectrum at low-$\ell$ multipole, and implies that $n_s$ decreases, $\epsilon$ and $r$ increase as $k_*$ goes to large scales.
Moreover, there could be
the acoustic wave of dark-matter density perturbation $\delta \rho_{_M}/\rho_{_M}$ in the ``pre-inflation'' epoch, 
described by the sound velocity $c^{M}_s=\omega_{_M}^{1/2}\not= 0$.
Analogously to baryon acoustic oscillations, these dark-matter sound waves should probably have imprinted in the both CMB 
and matter power spectra at large scales of $k_* \sim 10^{-3}{\rm Mpc}^{-1}$. 

\comment{
 However, we wish to see the way to theoretically determine the value $N_* =\ln (a_e/a_*)$, and understand the entropy issue in the reheating era. 
}

\noindent
{\bf Entropy and reheating.}
\comment{In order to see how to end the inflation, we need to discuss the entropy and particle creation. 
The entropy has two parts: 1. particles and 2. horizon $H$ 
If the reaction rate of $\Lambda$ or space-time 
$H$ creating matter and anti matter which annihilating to $\Lambda$ is faster than Universe expansion, we assume each of them 
is in thermal equilibrium with the characteristic temperature $T_M$ or $T_H$, where Hawking temperature $T_H=H/2\pi$ of De Sitter space. 
(may be Universe expansion is faster, than some anti-matter is out of horizon)}
The pair production from the spacetime is an entropically favorable process, and pairs can in turn annihilate back to the spacetime. Considering the processes as thermal emissions and absorptions, 
we discuss their entropy and temperatures associating 
with Eq.~(\ref{emt}), using the first law of thermal dynamics
in the volume $V=(4\pi/3) H^{-3}$,     
\begin{eqnarray}
dQ_M&=&T_MdS_M=d(\rho_{_M} V) +p_{_M}dV-\mu_{_M} dN,\label{enm}\\
dQ_\Lambda &=&T_\Lambda dS_\Lambda=d(\rho_{_\Lambda} V) +p_{_\Lambda} dV=d(\rho_{_\Lambda}) V,
\label{enL0}
\end{eqnarray}
where the entropy $S_\Lambda$ is related to the horizon entropy as shown below.
Since the chemical potentials of fermions $F$ and anti-fermions 
$\bar F$ are equal and opposite $\mu_{_M}=-\bar\mu_{_M}$ in the 
pair production $dN=d\bar N$, the chemical potential 
$\mu_{_M}=\mu_{_M}+\bar\mu_{_M}=0$ in Eq.~(\ref{enm}).  
From the particle number conservation 
$(n_{_M}U^a)_{;\,b}=0$ and 
the total energy conservation along the fluid flowing 
line $U_a(T^{ab}_{_M} + T^{ab}_{_\Lambda})_{;\,b}=0$, we obtain 
\begin{eqnarray}
T_MdS_M + T_\Lambda dS_\Lambda=0.
\label{tsc}
\end{eqnarray}
This relates to the total entropy conservation, implying the adiabatic evolution of the Universe composed by matter 
and spacetime. 
In fact, Eq.~(\ref{tsc}) is equivalent to the conservation law 
(\ref{fcgeqi20}), provided with Eqs.~(\ref{enm}) and (\ref{enL0}). However, the matter entropy $S_M$ renders the physical sense of 
the spacetime entropy $S_\Lambda$, via the back and forth processes of spacetime producing pairs, which annihilate to the spacetime. 

For no pair production $\Omega_{_M}=0$, 
$\Lambda = 3H^2$ and $H={\rm const.}$, Eq.~(\ref{enL0}) gives 
\begin{eqnarray}
S_\Lambda=\frac{\rho_{_\Lambda}}{T_\Lambda}V 
= \frac{3H}{4G} V 
=\frac{\pi M^2_{\rm pl}}{H^2}=\frac{1}{4}{\mathcal A}=S_H ,
\label{enL}
\end{eqnarray}
provided $T_\Lambda=T_H=H/2\pi$ Hawking temperature, where 
$S_H$ is the entropy of an De Sitter spacetime \cite{entropy}. 
The entropy (\ref{enL}) relates to the total number of spacetime 
quantum states ($q,p$) on the horizon area ${\mathcal A}=4\pi H^{-2}/\ell^2_{\rm pl}$, fluctuating $\delta q\cdot\delta p\approx 1$, 
$\delta q\approx \ell_{\rm pl}$ and $\delta p \approx M_{\rm pl}$ at the Planck scale. The characteristic state is
$\delta q \approx 2\pi H^{-1}$ and $\delta p \approx T_H$. 

\comment{$\rho_{_\Lambda}=\Lambda/(8\pi G)$ and $\rho_{_\Lambda}/T_H$ is the entropy density. 
where the comoving volume $V=(4\pi/3)H^{-3}$. 
Since $\rho_{_\Lambda}=-p_{_\Lambda}$, Eq.~(\ref{enL0}) becomes 
$T_HdS_\Lambda=d(\rho_{_\Lambda}) V$, where we can define the variation of entropy density $ds_\Lambda=d(\rho_{_\Lambda})/T_H$. Assuming $H$ does not change, then temperature and volume do not change, the total particle entropy 
$S_M=\rho_{_M} V/T_M$ and 
 The total entropy conservation requires $dS_M=-dS_\Lambda$ leading to $d\rho_{_M}=-d\rho_{_\Lambda}$. 
} 

In the inflation epoch $H>\Gamma_M$ 
and $\Omega_{_\Lambda}\gg \Omega_{_M}\not=0$,  
$H$ and $\Omega_{_\Lambda}$ slowly decrease, due to pair production. However the rate of pairs annihilating back to the spacetime is smaller than the inflation rate $H$, 
i.e., $\Gamma^{^{\rm anni}}_{M}= \Gamma_M < H$, 
so that the pairs 
are far from reaching an equilibrium or equipartition 
with the inflating spacetime. 
\comment{For $\Omega_{_M}\approx 0$, Eq.~(\ref{tsc}) leads to $S_\Lambda \approx S^\circ_\Lambda = {\rm const.}$, i.e.,
\begin{eqnarray}
S_\Lambda \approx \pi (T_\Lambda H^{-1})^2= \pi (T^\circ_\Lambda H^{-1}_\circ)^2, \quad
T^\circ_\Lambda =H_\circ/(2\pi),~~~ 
\label{enLf}
\end{eqnarray} 
As a result, the variations of $T_\Lambda$ and $H$ follow 
$T_\Lambda \approx H/(2\pi)$, and cosmological term 
$\rho_\Lambda  \approx 3(2\pi)^3T_\Lambda^4$.
} 
Equations (\ref{enL0}) and (\ref{tsc}) give
\begin{eqnarray}
T_MdS_M = - d(\rho_{_\Lambda})V \approx -d(3H^2/8\pi G) V 
= 2\,\epsilon\, T_{H}S_{H}dx,~~~ 
\label{tsc1}
\end{eqnarray} 
where $T_\Lambda\approx T_H \gg T_M$. 

After the inflation epoch, $\Gamma_M > H$ implies that 
pairs have large density and rate to annihilate back to the spacetime. 
This epoch should be studied by integrating 
Eqs.~(\ref{fe3}) and (\ref{fcgeqi20}) with 
the rate equation \cite{rate} 
\begin{eqnarray}
\frac{dn_{_M}}{dt}+ 3 Hn_{_M} = \Gamma_M \left(n_{_M}^{T_\Lambda} - n_{_M}\right), 
\label{rateeq}
\end{eqnarray}
where $n^{T_\Lambda}_{_M}$ is the thermal density 
of pairs in an equilibrium with the spacetime 
at the temperature $T_\Lambda$. If pairs reach a thermal equilibrium with the spacetime, namely $T_\Lambda =T_M=T_H$ 
and $n^{T_M}_{_M}=n_{_M}$ in Eq.~(\ref{rateeq}), $dn_{_M}/dx+3n_{_M}=0$ and $n_{_M}\propto a^{-3}$. The total entropy conservation 
$dS_M=-\,dS_\Lambda$ (\ref{tsc})
indicates that the spacetime entropy is converted to the matter entropy, as $\Omega_{_\Lambda}$ decreases and $\Omega_{_M}$ increases, 
\begin{eqnarray}
S_M(\tilde H) - S_M(H_e)=S_\Lambda (H_e)-S_\Lambda (\tilde H),
\label{centropy}
\end{eqnarray} 
from the inflation ending $H_e$ to the reheating 
end $\tilde H$. We may consider the approximations: (i) $S_M(H_e)\approx 0$ as $\Omega_{_M}(H_e)\ll \Omega_{_\Lambda}(H_e)$; (ii) 
$S_\Lambda (H_e)\approx S_{H}(H_e)$ of Eq.~(\ref{enL}) as 
$\rho_{_\Lambda}(H_e)\approx 3H_e^2$; (iii) the reheating ends 
up with $S_\Lambda (\tilde H) \approx 0$ and $\tilde H \approx 0$,  
so that the spacetime entropy converted to the matter entropy 
is maximal $S_M(\tilde H)\approx S_\Lambda (H_e)$. Actually, at a certain point the pairs decay to light particles rather than  annihilate to the spacetime, thus are out of thermal equilibrium, 
$n_{_M}^{T_\Lambda}$ exponentially decreases and $T_{M}> T_\Lambda$. The Universe stops acceleration and starts 
deceleration for $2\Omega_{_\Lambda}\leq (1+3\, \omega_{_M})\,\Omega_{_M}$.

\comment{leading to 
\begin{eqnarray}
dS_M=-\,dS_\Lambda=-\frac{d(\rho_{_\Lambda})}{T_\Lambda} V=f
d \left(\frac{\pi m_{\rm pl}^2}{H^2}\right) =f
d \left(\frac{\pi H^{-2}}{\ell_{\rm pl}^2}\right),
\label{enc}
\end{eqnarray}
where we assume $\Lambda=3fH^2$ and function 
$f=f(H/H_\circ)<1, f(1)=1$ weakly depends on the evolution 
$H/H_\circ$. Equation (\ref{enc}) shows 
that in the evolution the increasing matter entropy $dS_M$ is equal to the 
horizon area changes in unit of Planck area $\ell_{\rm pl}^2$ in the evolution.}

The enormous matter entropy (temperature) 
is generated (increased) by the decay
of massive pairs to light particles, when the decay rate 
$\Gamma^{\rm decay}_M \propto g^2_{_Y} m >\Gamma_M >H$, 
where $g_{_Y}$ is the Yukawa coupling between the massive pairs and
light particles. The term $\Gamma_M^{\rm decay}n_{_M}$ should be added to the RHS of Eq.~(\ref{rateeq}), and the particle number conservation law changes to  $(n_{_M}U^a)_{;b}=-\Gamma_M^{\rm decay}n_{_M}$. 
\comment{
the usual conservation law $d\Omega_{_M}/dx=-3(1+\omega_{_M})\Omega_{_M}$, i.e., is no longer valid, even in the case that particles decouple from the cosmological term. The decay processes can be preliminary described as
\begin{eqnarray}
d\Omega_{_M}/dx \approx -3 \,\omega^{\rm decay}_{_M}\Omega_{_M}, \quad \omega^{\rm decay}_{_M}\equiv \Gamma^{\rm decay}_M/(3H),
\label{decay}
\end{eqnarray}  
} 
Postponing detailed studies of the complex reheating epoch, 
we postulate  
the reheating epoch ends at $\tilde a$, $\tilde t$, 
$\tilde\rho_c=3\tilde H^2$, and $\tilde T $,  
when an enormous amount of light particles 
decouples from the  
$\Omega_{_\Lambda}$, and approximately follow their own conservation 
law ($x=\ln a/\tilde a$), 
\begin{eqnarray}
d\Omega_{_M}/dx \approx -3(1+\omega_{_M})\Omega_{_M},\quad \Omega_{_M}(\tilde a)=\tilde\Omega_{_M}\gg \tilde\Omega_{_\Lambda}.
\label{endrh}
\end{eqnarray}
We henceforth use $\Omega_{_M}$ and $\omega_{_M}$ to represent the 
``usual'' matter that had been produced by the end of the reheating, governing the Universe evolution later on.  

\comment{
As a result, $\Omega_{_\Lambda}$ almost decouples from 
Eq.~(\ref{fcgeqi20}), as if it had been frozen as a ``constant''. Nevertheless $\Omega_{_\Lambda}$ weakly coupling to $\Omega^{\rm l,h}_{_M}$ via $h$ and dominantly governs $h$ again, as decreasing 
$\Omega^{\rm l,h}_{_M}< \Omega_{_\Lambda}$. 
because $\Gamma \ll H$ and $T_M \gg T_H$ in this epoch, particles 
have no enough density and rate to annihilate back to the spacetime.
(iii) $\Gamma < H$, $T_M > T_H$. The space time decouples from matter particles, the latter evolves with this huge entropy in  standard cosmology, however there is still interaction between space time and matter particles, though they are decoupled each other from thermal 
equilibrium, and not in equilibrium. The particle matter density value at the heating era is given by Eq.~(\ref{nden}) at $H_{\rm end}$, 
and then follow the law $(a_{\rm end}/a)^4$ to decrease, assume they are a relativistic gas. }

\noindent
{\bf Cosmic coincidence.} \hskip0.01cm
In the standard cosmology epoch, we separately consider two 
matter contributions: (i) 
the ``coupled'' matter $\Omega^\Lambda_{_M}(h)$ ($\Omega^\Lambda_{_M}\ll \Omega_{_\Lambda}$) and $\omega^{\Lambda}_{_M}$, representing the
particle-antiparticle pairs produced {\it after} the reheating end, computed by Eqs.~(\ref{nden}-\ref{pden}) since $\tilde t=0$; 
(i) the usual matter $\Omega_{_M}\approx \tilde\Omega_{_M}\exp -3(1+\omega_{_M})\, x$ of Eq.~(\ref{endrh}), neglecting $\Omega_{_M}$-annihilation to $\Omega_{_\Lambda}$ and $\Omega^{\Lambda}_{_M}$-decay to $\Omega_{_M}$. In this approximation, the conservation law (\ref{fcgeqi20}) decouples 
into Eq.~(\ref{endrh}) and  
\begin{eqnarray}
\frac{d}{dx}\left(\Omega_{_\Lambda}+\Omega^\Lambda_{_M}\right)
\approx -3(1+\omega^{\Lambda}_{_M}+ \omega^{\rm decay}_{_M})\,\Omega^\Lambda_{_M},
\label{decay}
\end{eqnarray}
where we incorporate $(n_{_M}U^a)_{;\,b}=-\Gamma_M^{\rm decay}n_{_M}$ and introduce 
$\omega^{\rm decay}_{_M}\equiv \Gamma^{\rm decay}_M/H$ for 
particle-antiparticle pairs decay to 
relativistic/non-relativistic particles in the radiation/matter dominate epoch. The $\omega^{\rm decay}_{_M}$ value 
depends on the final states and phase space of particles that they subsequently decay. 
\comment{
The total entropy of the $\Omega_{_M}$,
$\Omega^{\Lambda}_{_M}$ and $\Omega_{_\Lambda}$ is conserved. 
Neglecting $\Omega_{_M}$-annihilation to $\Omega_{_\Lambda}$ 
and $\Omega^{\Lambda}_{_M}$-decay to $\Omega_{_M}$, 
it approximately decouples to the $\Omega_{_M}$-entropy 
conservation  (\ref{endrh}) and the entropy 
conservation (\ref{tsc}) of $\Omega_{_\Lambda}$ and $\Omega^{\Lambda}_{_M}$ previously discussed, however they all couple together 
via $H$.} 
As a result, Eqs.~(\ref{fe3}) and (\ref{fcgeqi20}) are recast as 
\begin{eqnarray}
h^2 = (\Omega_{_M} + \Omega_{_\Lambda}),\quad
\frac{d\Omega_{_\Lambda}}{dx}\approx  -3\,(1+
\omega^{\rm decay}_{_M})\Omega^\Lambda_{_M}(h),
\label{ggg}
\end{eqnarray}
where we rewrite $(\Omega^\Lambda_{_M} + \Omega_{_\Lambda})$ as a new
$\Omega_{_\Lambda}$, since it overall represents ``dark energy'' 
in observations. It shows $\Omega_{_\Lambda}$ 
indirectly interacting with 
$\Omega_{_M}$ through $h$. Note $\omega^{\Lambda}_{_M}\approx 0$ for $H/m\ll 1$, see Eq.~(\ref{apden}).
\comment{As a result, $\Omega_{_\Lambda}$ does not vanish. 
As an illustration in  Figure 2, we show numerical solutions to 
Eqs.~(\ref{rden},\ref{pden}) and (\ref{ffe3},\ref{ggg}). Figure caption  $\tilde H =10^{-8}$ and $m=M_{\rm pl}=(8\pi)^{1/2}\approx 5.01$.}

To calculate $\Omega^\Lambda_{_M}(h)$, we introduce another mass 
scale $\tilde m$ in Eq.~(\ref{rden}). 
In the physical regime $\tilde m \gg H$, using $\Omega^\Lambda_{_M}\approx 2 \chi \tilde m ^2 h^2/3$ in Eq.~(\ref{ggg}), analogously to the asymptotic expressions (\ref{aden},\ref{apden}) for $\chi m^2\rightarrow\chi\tilde m^2\ll 1$, 
we obtain
\begin{eqnarray}
\frac{d\Omega_{_\Lambda}}{dx} +\tau\Omega_{_\Lambda}=-\tau\,\Omega_{_M},\quad \tau\equiv 2\,\chi\, \tilde m^2\,(1+\omega^{\rm decay}_{_M}).
\label{gg1}
\end{eqnarray}
In the radiation dominate epoch 
starting from the reheating 
end (\ref{endrh}), the solution ($x=\ln a/\tilde a$ and $\omega_{_M}=1/3$) is
\begin{eqnarray}
\Omega_{_\Lambda} = \frac{\tau_{_R} \tilde\Omega_{_M}}{4-\tau_{_R}}e^{-4x} +e^{-\tau_{_R}x}\,\tilde{\mathcal C}=\frac{\tau_{_R}}{4-\tau_{_R}}\Omega_{_M} \propto h^2,
\label{gg2}
\end{eqnarray}
where $\tau_{_R}\approx 2\,\chi\, \tilde m^2\,[1+\omega^{\rm decay}_{_{M,R}}]$. 
Here we 
choose the initial condition 
$\tilde {\mathcal C}=0$ at $a=\tilde a$, 
i.e., $\tilde \Omega_{_\Lambda}=\tau_{_R}\tilde
\Omega_{_M}/(4-\tau_{_R})$ at the reheating end (\ref{endrh}), 
for the reason that the transitions from the reheating end to 
the standard cosmology start are radiation dominate and continuous, they have 
the same $\omega_{_M}$ and 
$\omega^{\rm decay}_{_{M,R}}$ values. 
Solution (\ref{gg2}) shows that in a long dark epoch,  
$\Omega_{_\Lambda}\ll \Omega_{_M}$ and $\Omega_{_\Lambda}$ tracks \cite{track} down 
$\Omega_{_M}$ 
until the Universe reaches the radiation-matter equilibrium 
$(a_{\rm eq}/\tilde a) = (\tilde  T/T_{\rm eq})\sim 10^{15}{\rm GeV}/ 10\,{\rm eV}\sim 10^{23}$, 
\begin{eqnarray}
\Omega^{\rm eq}_{_\Lambda}\approx 
(\tau_{_R}/4)\,\Omega^{\rm eq}_{_M}\ll 1,\quad \Omega^{\rm eq}_{_M}
=\Omega_{_M}(a_{\rm eq})\lesssim 1,
\label{equi}
\end{eqnarray}
in unit of the density $\rho_c^{\,\rm eq}=3 H^2_{\rm eq}$. Equations (\ref{fe3}) and (\ref{gg2}) give $\Omega_{_\Lambda}\approx (\tau_{_R}/4)\, h^2$ in this dark epoch.  
\comment{here we see that in this long dark age of Universe, the $\Omega_{_\Lambda}$ is much smaller than the total matter components, and also had been eventually suppressed to the negligible value, even though the value at the reheating end could be large, provided $\chi \tilde m^2\sim 10^{-11}$ is small enough for $m\sim 10^{-4}$, see below. If $m\sim 10^{-4}$ is this value, the suppression is not very large, what is account is the value $\tilde\Omega_{_\Lambda}\approx 0$ at the reheating end.}
 
In the matter dominate epoch starting from $a_{\rm eq}$ (\ref{equi}) to the present epoch $a\simeq a_0$ and $(a/a_{\rm eq})\simeq (1+z)\sim 10^4$, the solution to Eq.~(\ref{gg1}) ($x=\ln a/a_{\rm eq}$ 
and $\omega_{_M}=0$) is
\begin{eqnarray}
\Omega_{_\Lambda} 
=\frac{\tau_{_M}}{3-\tau_{_M}}\Omega_{_M} + e^{-\tau_{_M}\,x}{\mathcal C}^{\rm eq}\approx \frac{\tau_{_M}}{3}\,h^2 + {\mathcal C}^{\rm eq},~~ 
\label{gg3}
\end{eqnarray}
where $\tau_{_M}\approx 2\,\chi\, \tilde m^2\,[1+\omega^{\rm decay}_{_{M,M}}]$ and the initial condition ${\mathcal C}^{\rm eq}$ is fixed by 
Eq.~(\ref{equi})
\begin{eqnarray}
{\mathcal C}^{\rm eq}=
2\,\chi\tilde m^2\,\Delta\omega^{\rm decay}_{_M}\,\Omega^{\rm eq}_{_M}. 
\label{ceq}
\end{eqnarray}
The $\Delta\omega^{\rm decay}_{_M}$ represents the 
effective variation from $\omega^{\rm decay}_{_{M,R}}$  to 
$\omega^{\rm decay}_{_{M,M}}$, and $\Delta\omega^{\rm decay}_{_M}>0$ for a larger and recursively generated phase space of final states of particles and their subsequent decays \cite{pdgdecay}. 
Actually, the ${\mathcal C}^{\rm eq}$ (\ref{ceq}) is the integration over discontinuous transitions from the radiation dominate epoch 
to the matter dominate one. 

\begin{figure}   
\includegraphics[height=7.0cm,width=12.0cm]{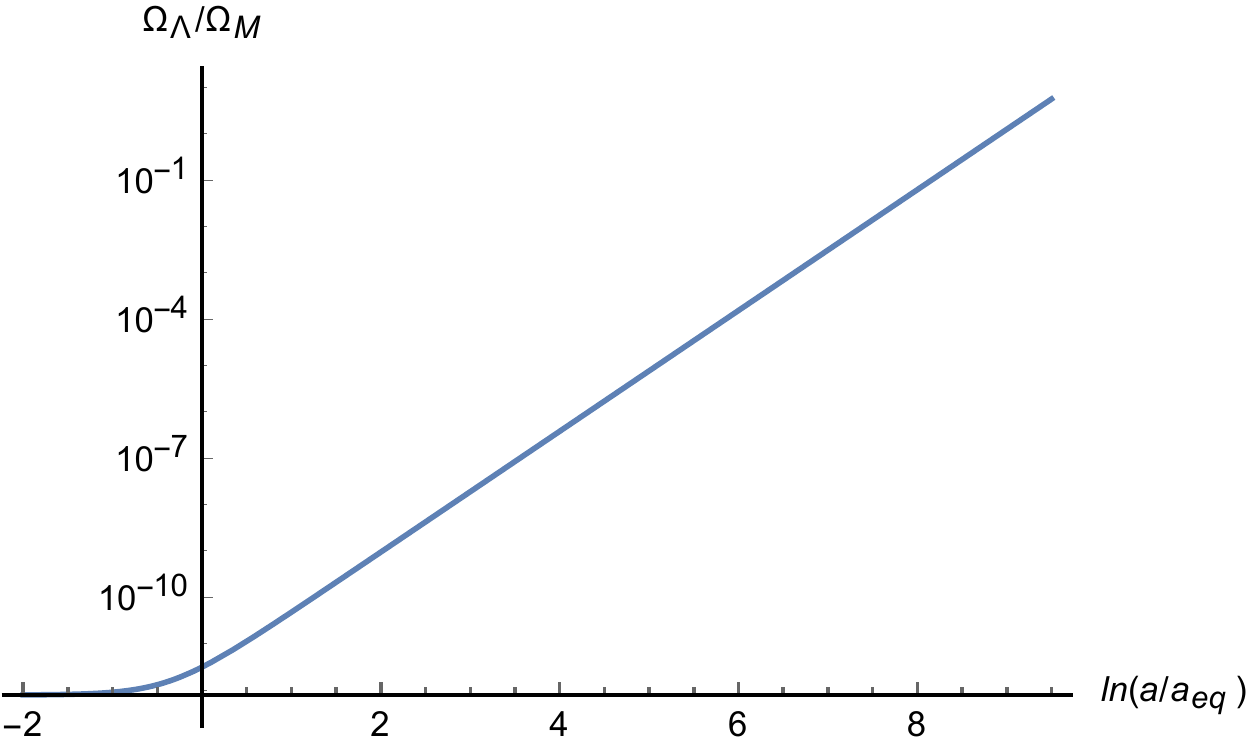}
\vspace{-1em}
\caption{The ratio $\Omega_{_\Lambda}/\Omega_{_M}$ (\ref{gg5}) is plotted as 
a function of $\ln (a/a_{\rm eq})$, where the scaling factor $a$ 
runs from the reheating end $\tilde a$, through the radiation-matter equilibrium 
$a_{\rm eq}$ to the present time $a_0$, $\tilde a<a_{\rm eq}<a_0$. 
It shows that (i) the tracking-down 
behavior: the ratio is a small constant $\sim 10^{-12}$ for 
$\ln (a/a_{\rm eq})<0$;  (ii) the tracking-down failure occurs around
the radiation-matter equilibrium $\ln (a/a_{\rm eq})=0$; (iii) 
$\Omega_{_\Lambda}\approx {\rm const.}$  
and $\Omega_{_M}\sim (a/a_{\rm eq})^{-3}$, the ratio 
$\Omega_{_\Lambda}/\Omega_{_M}$ increases to 
${\mathcal O}(1)$
at the present time $(a/a_{\rm eq})\sim 10^4$ and $\ln (a/a_{\rm eq})\approx 9.2$. 
When $\Omega_{_\Lambda}/\Omega_{_M}=1/2$, the Universe turns from 
the deceleration phase to the acceleration phase. The cosmological term 
$\Omega_{_\Lambda}$ will dominate over the matter term $\Omega_{_M}$ in future.}
\label{ccplot}
\end{figure}

In this light epoch,
Eq.~(\ref{gg3}) shows that the first term decreases as $\Omega_{_M}\approx \,
\Omega^{\rm eq}_{_M}(1+z)^{-3}$ and
$\Omega_{_\Lambda}$ fails to track down $\Omega_{_M}$, approaching to a slowly varying ``constant'' $e^{-\tau_{_M}\,x}\,{\mathcal C}^{\rm eq}$, which recalls its value (\ref{equi}) 
at the radiation-matter equilibrium. 
As a result, we obtain the ratio 
\begin{eqnarray}
\Omega_{_\Lambda}/\Omega_{_M}
\approx (\tau_{_M}/3) + 2\,\chi\tilde m^2\,\Delta\omega^{\rm decay}_{_M}\,(1+z)^{3},\quad (1+z)=(a/a_{\rm eq})^{-1}
\label{gg5}
\end{eqnarray}
Using current observations $\Omega^0_{_\Lambda}\approx 0.7$ and 
$\Omega^{0}_{_M}\approx 0.3$, correspondingly $z\sim 10^4$, we obtain 
$2\chi\tilde m^2 \Delta\omega^{\rm decay}_{_M}\approx  (1+z)^{-3}
\Omega_{_\Lambda}/\Omega_{_M}  \approx 2.3\times 
10^{-12}$. If  
$\Delta\omega^{\rm decay}_{_M}\sim {\mathcal O}(1)$, 
$\tau_{_M}\approx \tau_{_R} \sim {\mathcal O}(10^{-12})$ and 
the mass scale $\tilde m \sim 10^{14}$ GeV 
coincides with the reheating temperature $\tilde T$. \comment{This result seems natural without any fine tuning, since the light epoch of $z\sim  10^{3\sim 4}$ is much shorter than the dark epoch of $(a^{\rm eq}/\tilde a)\sim 10^{23}$, when 
the $\Omega_{_\Lambda}$ tracks down $\Omega_{_M}$. This gives us an insight into the cosmic coincidence. Otherwise we would have the cosmic coincidence problem
of an incredibly fine tuning $\sim (10^{23})^4\times (10^4)^3$ 
of the values $\tilde \Omega_{_\Lambda}$ and $\tilde \Omega_{_M}$ 
at the reheating end so as to reach their present observational values 
of the same order of magnitude.} 
These results give us an insight into the issue of the cosmic coincidence 
at the present time.  
The $\Omega_{_\Lambda}$ and $\Omega_{_M}$ relation shows that 
the cosmic coincidence of $\Omega_{_\Lambda}$ and 
$\Omega_{_M}$ values appears naturally without any extremely fine tuning, 
since the matter dominated epoch of $z\sim  10^{3\sim 4}$ 
is much shorter than the radiation dominated epoch 
of $(a^{\rm eq}/\tilde a)\sim 10^{23}$, when 
the $\Omega_{_\Lambda}$ tracks down $\Omega_{_M}$ and the ratio 
$\Omega_{_\Lambda}/\Omega_{_M}$ is constant.  
Otherwise we would have the cosmic coincidence problem
of an incredibly fine tuning the values $\tilde \Omega_{_\Lambda}$ 
and $\tilde \Omega_{_M}$ 
at the reheating end at the order 
$\sim (10^{-23})^4\times (10^{-4})^3\sim 10^{-104}$, 
so as to reach their present observational values 
of the same order of magnitude. To describe this scenario, 
we use the ratio $\Omega_{_\Lambda}/\Omega_{_M}$, 
which is independent of the different 
units used in different epochs. In Fig.~\ref{ccplot}, we plot the ratio 
$\Omega_{_\Lambda}/\Omega_{_M}$ from the radiation dominated epoch (\ref{gg2})
to the matter dominated epoch (\ref{gg5}) for an explicit illustration.

The $\Omega_{_\Lambda}-\Omega_{_M}$ relation (\ref{gg3}) can be rewritten in units of the critical density $\rho_c=3H_0^2$ today,
\begin{eqnarray}
\Omega_{_\Lambda} \approx (\tau_{_M}/3)\,\Omega^0_{_M}(1+z)^3 + \Omega^0_{_\Lambda} (1+z)^{\tau_{_M}},
\label{gg6}
\end{eqnarray}
and $\Omega^0_{_\Lambda}\approx \Omega^{\rm eq}_{_\Lambda}\sim 10^{-12} \,\Omega^{\rm eq}_{_M}$.  
This can possibly be examined with observations \cite{clement}. 
In particular, how to examine the 
$\Omega_{_\Lambda}$-transition 
from the present ``constant'' $\sim (1+z)^{\tau_{_M}}$ back 
to the track-down evolution $\sim (1+z)^3$ at the large 
redshift $z\sim 10^{3\sim 4}$. We speculate that such 
$\Omega_{_\Lambda}$-transition 
should induce the peculiar fluctuations of gravitational field that imprint on the CMB spectrum, analogously to the integrated 
Sachs-Wolfe effect.   


\noindent
{\bf Some remarks.}
We emphasize that the area law (\ref{aden}) and (\ref{apden}) are crucial for obtaining the law $\Omega_{_\Lambda}\propto h^2$ in the cosmic inflation 
and $\Omega_{_\Lambda}\sim \Omega_{_M}$ coincidence in the present time. 
The initial value $\Omega^\circ_{_\Lambda}\propto H^2_\circ$ at the 
Planck scale should be attributed to the spacetime quantum fluctuation 
at the Planck scale \footnote{For some more discussions, see Refs.~\cite{cosh,xueNPB2015}.}. 
Oppositely to the matter and its negative gravitational potential, 
the $\Omega_{_\Lambda}$ physically
represents a negative mass, whose positive potential not only leads to the pair production, but also to the Universe acceleration. 
In turn, these pairs ``screen'' the positive potential, 
increase $\Omega_{_M}$ and deepen the negative potential. 
The present value 
$\Omega^0_{_\Lambda}\propto H_0^2$ 
\footnote{The ``area'' law of 
``vacuum-energy'' density  
$\rho^{\rm vac}_{_\Lambda}\approx \pi/(2\ell^2_{\rm pl} H_0^{-2})$, 
rather than 
$\rho^{\rm vac}_{_\Lambda}\propto 1/(\ell^4_{\rm pl})$, 
V.~G.~Gurzadyan and S.-S.~Xue, IJMPA 18 (2003) 561-568, 
astro-ph/0105245,  
and more discussions, see S.-S.~Xue, IJMPA 24 (2009) 3865-3891, arXiv:hep-th/0608220 . } is 
the consequence of $\Omega_{_\Lambda}$ creating and interacting with $\Omega_{_M}$ in the Universe evolution. 
The positivity of total energy is expected as long as Eqs.~(\ref{fe3})and (\ref{fcgeqi20}) hold. 
Full numerical approach to this problem is very inviting.  The lengthy article in details can be found in Ref. \cite{xue2020}.

Author thanks Dr.~Yu Wang for an indispensable numerical assistance of using Python. 

\end{document}